\documentclass[shortnote,twocolumn]{jpsj2}

\newcommand{\nc}{\newcommand}
\nc{\rnc}{\renewcommand}
\nc{\nn}{\nonumber}
\rnc{\i}{{\rm i}}
\rnc{\d}{{\rm d}}
\nc{\e}{{\rm e}}
\nc{\bra}{\langle}
\nc{\ket}{\rangle}
\nc{\bbra}{\left\langle}
\nc{\kket}{\right\rangle}
\rnc{\(}{\left(}
\rnc{\)}{\right)}
\rnc{\[}{\left[}
\rnc{\]}{\right]}

\title{On the density matrix for the kink ground state of higher spin XXZ chain}

\author{Kohei Motegi}

\inst{Institute of Physics, University of Tokyo, Meguro-ku, Tokyo 153}

\kword{XXZ chain, kink ground state, correlation function}

\begin{document}
\maketitle

The exact computation of the correlation functions
of 1D quantum integrable models
has been one of the challenging problems.
For the spin-1/2 XXZ chain, the correlation functions
in the {\it antiferromagnetic} regime were found to
be expressed in the multiple integral form.$^{1,2)}$
However, the exact evaluation of them is still a hard work,
and has been only successful for some special 
anisotropy parameters.$^{3)}$

In the {\it ferromagnetic} regime, there is a class 
of nontranslationally invariant ground state
which should be called as {\it kink} ground state$^{4,5)}$.
We have studied the correlation function of the {\it kink}
ground state, and exactly calculated the density matrix$^{6)}$
(see also ref 7).

The kink ground state also exists for arbitrary spin and dimension.
In this paper, we extend the analysis developed in our previous paper
to higher spin 1D XXZ chain, and calculate the density matrix
(note that the model we consider in this paper
is not the {\it integrable} higher spin XXZ chain$^{8)}$.

The Hamiltonian of the spin $S$ 1D infinite XXZ chain is
\begin{equation}
H=- \sum_{m \in \mathbb{Z}}
( S_{m}^x S_{m+1}^x+S_{m}^y S_{m+1}^y
+\Delta( S_{m}^z S_{m+1}^z-S^2) ).
\end{equation}
We shall consider 
the ferromagnetic regime $\Delta>1$. For later convenience,
we parametrize $\Delta$ as
$
\Delta=(q^{\frac{1}{2}}+q^{-\frac{1}{2}})/2.
$
Then $\Delta>1$ corresponds to $0<q<1$.
A kink ground state is the superposition
of kinks which have the same center.
For a (normalized) kink state $\otimes_{x \in \mathbb{Z}} |
m_x \ket$ $(m_x \in \{-S, \cdots ,S \})$,
the center $j-1/2 \ (j \in \mathbb{Z})$ 
is the position where
\begin{equation}
\sum_{x < j}(S-m_x)= \sum_{x >j}(m_x+S), \nn
\end{equation}
holds. 
Let us denote the kink ground state whose center is 
at $j-\frac{1}{2}$ by $| \Psi_j \ket$,
and introduce the generating function
of $| \Psi_j \ket^{9)}$
\begin{align}
| \Psi(z) \ket=&
  \bigotimes_{x \in \mathbb{Z}_{<0}} 
   ( \sum_{m_x=-S}^S (zq^{\frac{1}{2}(\frac{1}{2}+x)})^{m_x-S}
   c(m_x)^{\frac{1}{2}}  | m_x  \ket)                         \nn \\
&\otimes
  \bigotimes_{y \in \mathbb{Z}_{\geq 0}} 
   ( \sum_{m_y=-S}^S (zq^{\frac{1}{2}(\frac{1}{2}+y)})^{m_y+S}
   c(m_y)^{\frac{1}{2}}
    | m_y  \ket),
\label{GF}
\end{align}
where $c(m_x)=(2S)!/(S-m_x)!(S+m_x)!$.
$|\Psi_j \ket$ is the coefficient of $z^{j}$ of the expansion of
$| \Psi(z) \ket$, i.e,
\begin{equation}
| \Psi(z) \ket=\sum_{j \in \mathbb{Z}} z^j |\Psi_j \ket.
\end{equation}
Let us focus on one of the kink ground states
$| \Psi_0 \ket$, and calculate the density matrix
\begin{equation}
\bra \prod_{j=1}^n 
E_{x_j}^{\epsilon_j^{\prime} \epsilon_j} \ket
:=
\frac{\bra \Psi_0 |   
\prod_{j=1}^n 
E_{x_j}^{\epsilon_j^{\prime} \epsilon_j}
| \Psi_0 \ket }
{\bra \Psi_0 | \Psi_0 \ket},
\end{equation}
where $E_x^{\epsilon_x^{\prime} \epsilon_x}| m_x \ket=
\delta_{m_x, \epsilon_x}| \epsilon_x^{\prime} \ket$.
$x_j$ is the position of the site where the operator 
$E_{x_j}^{\epsilon_j^{\prime} \epsilon_j}$ acts on, and 
is  assumed to be $x_j \neq  x_k$ for $j \neq k$.
We only consider the case when 
$\sum_{j=1}^n(\epsilon_j-\epsilon_j^{\prime})=0$
is satisfied: otherwise, the density matrix is zero.
We calculate $\bra  \Psi(z) | \Psi(z)  \ket  $
and $\bra  \Psi(z) |   \prod_{j=1}^n 
E_{x_j}^{\epsilon_j^{\prime} \epsilon_j}
| \Psi(z)  \ket$ to obtain the exact expression
of the density matrix since
\begin{align}
\bra \Psi(z) | \Psi(z) \ket &= \sum_{j=-\infty}^{\infty} z^{2j}
\bra \Psi_j | \Psi_j \ket, \\
\bra \Psi(z) |
\prod_{j=1}^n E_{x_j}^{\epsilon_j^{\prime} \epsilon_j}
| \Psi(z) \ket &=\sum_{j=-\infty}^{\infty} z^{2j} \bra \Psi_{j} |
\prod_{j=1}^n E_{x_j}^{\epsilon_j^{\prime} \epsilon_j} |
\Psi_{j} \ket,
\end{align}
holds,  and $\bra \Psi_0 | \Psi_0 \ket$, 
$\bra \Psi_0 |   
\prod_{j=1}^n 
E_{x_j}^{\epsilon_j^{\prime} \epsilon_j}
| \Psi_0 \ket $
can be extracted from $\bra  \Psi(z) | \Psi(z)  \ket  $ and
$\bra  \Psi(z) |   
\prod_{j=1}^n 
E_{x_j}^{\epsilon_j^{\prime} \epsilon_j}
| \Psi(z)  \ket$, which are easier to calculate.
Let us first calculate  $\bra  \Psi(z) | \Psi(z)  \ket  $.
Using the Jacobi triplet product identity
\begin{eqnarray}
(q;q)_{\infty} (-xq^{\frac{1}{2}};q)_{\infty}
(-x^{-1}q^{\frac{1}{2}};q)_{\infty}
=\sum_{j=-\infty}^{\infty}x^j q^{\frac{j^2}{2}}, \label{JT}
\end{eqnarray}
we have
\begin{align}
\bra \Psi(z) | \Psi(z) \ket=&(-wq^{\frac{1}{2}};q)_{\infty}^{2S}
(-w^{-1}q^{\frac{1}{2}};q)_{\infty}^{2S} \nn \\
=& \frac{1}{(q;q)_{\infty}^{2S}}
\Big( \sum_{j=-\infty}^{\infty}w^j q^{\frac{j^2}{2}} \Big)^{2S} \nn \\
=&\frac{1}{(q;q)_{\infty}^{2S}} \sum_{j=-\infty}^{\infty} A_j w^j,
\label{comp1}
\end{align}
where $w=z^2$, $(a;q)_{\infty}:=\prod_{j=0}^{\infty}(1-a q^{j})$
and
\begin{equation}
A_j=\sum_{\sum j_k =j, j_k \in \mathbb{Z}} \prod_{k=1}^{2S} q^{\frac{j_k^2}{2}}.
\nn
\end{equation}
Using eq. \eqref{JT}
and the following identity$^{6)}$
\begin{equation}
\prod_{j=1}^n \frac{1}{1+x u_j}=
\sum_{j=0}^{\infty} (-x)^j \sum_{l=1}^n 
\frac{u_l^{j+n-1}}{\prod_{i \neq l}(u_l-u_i)}, \label{identity}
\end{equation}
$\bra  \Psi(z) |   
\prod_{j=1}^n 
E_{x_j}^{\epsilon_j^{\prime} \epsilon_j}
| \Psi(z)  \ket$ can be calculated as
\begin{align}
&\bra \Psi(z) | \prod_{j=1}^n E_{x_j}^{\epsilon_j^{\prime} \epsilon_j} | \Psi(z) \ket
\nn \\ 
&=
\prod_{j=1}^n 
\{ (w \zeta_j)^{\[ \epsilon_j^{\prime} \epsilon_j \]}
c(\epsilon_j) c(\epsilon_j^{\prime}) \}
\prod_{j=1}^n \frac{1}{(1+w \zeta_j)^{2S}} \nn \\
&\times 
(-wq^{\frac{1}{2}};q)_{\infty}^{2S}
(-w^{-1}q^{\frac{1}{2}};q)_{\infty}^{2S} \nn \\
&=\frac{1}{(q;q)_{\infty}^{2S}}
\prod_{j=1}^n 
\{ (w \zeta_j)^{\[ \epsilon_j^{\prime} \epsilon_j \]}
c(\epsilon_j) c(\epsilon_j^{\prime}) \} \nn \\
&\times \sum_{j=-\infty}^{\infty} \Big(
\sum_{\substack{j_1 \in \mathbb{Z},
j_2 \in \mathbb{Z}_{\ge 0} \\ j_1+j_2=j}} A_{j_1} B_{j_2}
\Big) w^j,
\label{comp2}
\end{align}
where $\zeta_j=q^{\frac{1}{2}+x_j}$, 
$\[ \epsilon_j^{\prime} \epsilon_j  \]
=(\epsilon_j+\epsilon_j^{\prime}+2S)/2$ and
\begin{equation}
B_j=\sum_{\sum j_k =j, j_k \in \mathbb{Z}_{\ge 0}} 
\prod_{k=1}^{2S} \left\{
(-1)^{j_k} \sum_{l=1}^n 
\frac{\zeta_l^{j_k+n-1}}{\prod_{i \neq l}(\zeta_l-\zeta_i)}
\right\}. \nn
\end{equation}
From eqs. \eqref{comp1} and \eqref{comp2}, one has
\begin{align}
\bra \Psi_0 | \Psi_0 \ket &=\frac{A_0}{(q;q)_{\infty}^{2S}}, \\
\bra \Psi_0 | \prod_{j=1}^n E_{x_j}^{\epsilon_j^{\prime} \epsilon_j}
| \Psi_0 \ket &=\frac{1}{(q;q)_{\infty}^{2S}}
\prod_{j=1}^n 
\{ \zeta_j^{\[ \epsilon_j^{\prime} \epsilon_j \]}
c(\epsilon_j) c(\epsilon_j^{\prime}) \} \nn \\
&\times
\sum_{\substack{j_1 \in \mathbb{Z},
j_2 \in \mathbb{Z}_{\ge 0} \\ j_1+j_2=
-\sum_{j=1}^n \[ \epsilon_j^{\prime} \epsilon_j  \]
}} A_{j_1} B_{j_2},
\end{align}
leading to the exact expression of the density matrix
\begin{align}
\bra \prod_{j=1}^n 
E_{x_j}^{\epsilon_j^{\prime} \epsilon_j} \ket
&=\prod_{j=1}^n 
\{ \zeta_j^{\[ \epsilon_j^{\prime} \epsilon_j \]}
c(\epsilon_j) c(\epsilon_j^{\prime}) \} \nn \\
&\times
\sum_{\substack{j_1 \in \mathbb{Z},
j_2 \in \mathbb{Z}_{\ge 0} \\ j_1+j_2=
-\sum_{j=1}^n \[ \epsilon_j^{\prime} \epsilon_j  \]
}} \frac{A_{j_1} B_{j_2}}{A_0}.
\label{density1}
\end{align}
One can check that for $S=1/2$, eq. \eqref{density1}
recovers the result in ref. 6. However,
the expression of eq. \eqref{density1} gets more complicated 
as the spin becomes higher.

Concentrating on the $S=1$ case, 
we can derive some slightly easier expression by using
\begin{align}
\prod_{j=1}^n \frac{1}{(1+x u_j)^2}&=
\sum_{j=0}^{\infty} (-x)^j X_{n,j}, \nn \\
X_{n,j}&=\lim_{u_{i+n} \to u_i, i=1, \cdots n}
\sum_{l=1}^{2n}
\frac{u_l^{j+2n-1}}{\prod_{i \neq l}(u_l-u_i)}, \nn
\end{align}
which is a special case of eq. \eqref{identity}. The result is
\begin{align}
&\bra  \prod_{j=1}^n E_{x_j}^{\epsilon_j^{\prime} \epsilon_j} \ket
=\prod_{j=1}^n 
\{ \zeta_j^{\[ \epsilon_j^{\prime} \epsilon_j \]}
c(\epsilon_j) c(\epsilon_j^{\prime}) \} \nn \\
&\times \sum_{j=0}^{\infty} (-1)^j X_{n,j}
q^{\frac{1}{4}(j+\sum_{k=1}^n \[ \epsilon_k^{\prime} \epsilon_k \])^2} \nn \\
&\times (\delta_{j+\sum_{k=1}^n \[ \epsilon_k^{\prime} \epsilon_k \]}^{even}
+C \delta_{j+\sum_{k=1}^n \[ \epsilon_k^{\prime} \epsilon_k \]}^{odd}),
\label{density2}
\end{align}
where 
$C=2 \sum_{k=1}^{\infty} 
q^{(k-\frac{1}{2})^2}/(1+2 \sum_{k=1}^{\infty} q^{k^2})$.
From eq. \eqref{density2}, the magnetization and the spin-spin correlation functions
can be easily calculated.
\begin{align}
\bra S_x^z \ket&= \sum_{j=0}^{\infty} (-1)^j (j+1) \zeta^j 
q^{\frac{j^2}{4}} \nn \\
&\times (\zeta^2 q^{j+1}-1)
(\delta_{j}^{even}+C \delta_{j}^{odd}), \\
\bra S_{x_1}^z S_{x_2}^z \ket&=
\sum_{j=0}^{\infty} (-1)^j X_{2,j} (\delta_{j}^{even}+C \delta_{j}^{odd})
\nn \\
& \times (\zeta_1^2 \zeta_2^2 q^{\frac{(j+4)^2}{4}}+q^{\frac{j^2}{4}}
-(\zeta_1^2 +\zeta_2^2) q^{\frac{(j+2)^2}{4}}), \\
\bra S_{x_1}^+ S_{x_2}^- \ket&=
4 \zeta_1^{\frac{1}{2}} \zeta_2^{\frac{1}{2}} \sum_{j=0}^{\infty}(-1)^j X_{2,j}
\nn \\
& \times \{
(\delta_{j}^{even}+C \delta_{j}^{odd})(\zeta_1+\zeta_2) q^{\frac{(j+2)^2}{4}} \nn \\
&+(\delta_{j}^{odd}+C \delta_{j}^{even})(q^{\frac{(j+1)^2}{4}}+\zeta_1 \zeta_2
q^{\frac{(j+3)^2}{4}}) \},
\end{align}
where
\begin{align}
X_{2,j}=\frac{(j+1)(\zeta_1^{j+3}-\zeta_2^{j+3})
-(j+3) \zeta_1 \zeta_2 (\zeta_1^{j+1}-\zeta_2^{j+1})}{(\zeta_1-\zeta_2)^3}.
\nn
\end{align}
From these expressions, one can easily
show that the spin-spin correlation functions decay
exponentially for large distances.
\begin{align}
&\bra S_{x_1}^z S_{x_2}^z \ket- \bra S_{x_1}^z \ket
\bra S_{x_2}^z \ket \sim 
A^{zz}(x_1) q^{x_2+\frac{1}{2}} \text{ for $x_2 \gg 1$},  \\
&A^{zz}(x_1)=
2 \sum_{j=0}^{\infty} (-1)^j q^{\frac{j^2}{4}}(1- \zeta_1^2 q^{j+1}) \nn \\
&\times (j \zeta_1^{j-1}+(j+1)q^{\frac{1}{4}}C \zeta_1^j)
(\delta_{j}^{even}+C \delta_{j}^{odd}), \nn \\
&\bra S_{x_1}^+ S_{x_2}^- \ket \sim
A^{+-}(x_1)q^{\frac{1}{2}(x_2+\frac{1}{2})}
\text{ for $x_2 \gg 1$},  \\
&A^{+-}(x_1)=4 \zeta_1^{\frac{1}{2}}
\sum_{j=0}^{\infty} (-\zeta_1)^j (j+1) \nn \\
&\times \{
(\delta_{j}^{even}+C \delta_{j}^{odd}) \zeta_1 q^{\frac{(j+2)^2}{4}}
+(\delta_{j}^{odd}+C \delta_{j}^{even}) q^{\frac{(j+1)^2}{4}}
\}. \nn
\end{align}

\section*{Acknowledgment}
The author thanks K. Sakai for useful discussions and comments
on this work. This work  was partially  
supported by Global COE Program
(Global Center of Excellence for Physical Sciences Frontier)
from MEXT, Japan.

\label{lastpage}

\end{document}